\documentclass[preprint,review,12pt]{elsarticle}

\usepackage{amssymb}
\usepackage{amsmath}

\usepackage{bm}
\usepackage{mathtools}
\usepackage{caption}
\usepackage{graphicx}
\usepackage{latexsym}
\usepackage{booktabs}
\usepackage{multirow}
\usepackage{algorithm}
\usepackage{algpseudocode}
\usepackage{physics}
\usepackage{xcolor}
\usepackage{pgf}
\usepackage{braket}
\usepackage{autobreak}
\usepackage{makecell}
\usepackage{verbatim}
\usepackage{geometry}
\usepackage{subcaption}
\usepackage{hyperref}\hypersetup{hidelinks}
\usepackage{cleveref} 
\graphicspath{{Fig/}}
\graphicspath{{Fig/template/}}

\newcommand{\etal}{\textit{et al}.}

\newcommand{\lrr}[1]{\left(#1\right)}
\newcommand{\lrs}[1]{\left[#1\right]}

\usepackage{lineno}

\journal{}

\biboptions{sort&compress}

\begin{document}

\begin{frontmatter}




\title{Modeling Noise in Quantum Computing of Scalar Convection}



\author[fir]{Jiahua Yang}

\author[fir]{Zhen Lu\corref{cor1}}\ead{zhen.lu@pku.edu.cn}

\author[fir,sec]{Yue Yang\corref{cor1}}\ead{yyg@pku.edu.cn}

\cortext[cor1]{Corresponding author.}

\address[fir]{State Key Laboratory for Turbulence and Complex Systems, School of Mechanics and Engineering Science, Peking University, Beijing 100871, China}
\address[sec]{HEDPS-CAPT, Peking University, Beijing 100871, China}

\begin{abstract}
Quantum computing holds potential for accelerating the simulation of fluid dynamics. 
However, hardware noise in the noisy intermediate-scale quantum era significantly distorts simulation accuracy. 
Although error magnitudes are frequently quantified, the specific physical effects of quantum noise on flow simulation results remain largely uncharacterized. 
We investigate the influence of gate noise on the quantum simulation of one-dimensional scalar convection. 
By employing a quantum spectral algorithm where ideal time advancement affects only Fourier phases, we isolate and analyze noise-induced artifacts in spectral magnitudes. 
We derive a theoretical transition matrix based on Hamming distances between computational basis states to predict spectral decay, and then validate this model against density-matrix simulations and experiments on a superconducting quantum processor. 
Furthermore, using data-driven sparse regression, we demonstrate that quantum noise manifests in the effective partial differential equation primarily as artificial diffusion and nonlinear source terms. 
These findings suggest that quantum errors can be modeled as deterministic physical terms rather than purely stochastic perturbations.
\end{abstract}


\end{frontmatter}

\newpage

\section{Introduction}

High-fidelity computational fluid dynamics, particularly direct numerical simulation, faces a fundamental bottleneck due to the prohibitive computational costs~\cite{Pope2000,Moin2024Fundamentals}. 
This limitation motivates the exploration of quantum computing as a disruptive paradigm~\cite{Daley2022Practical,Hoefler2023,Waintal2024quantum}. 
By leveraging quantum properties such as superposition and entanglement, 
quantum algorithms promise theoretical exponential speedup for specific problems~\cite{nielsen2010}. 
Notable examples include algorithms for solving linear algebraic systems and differential equations~\cite{Harrow2009,Childs2017LCU,Berry2017Quantum,Gilyen2019QSVT,Subasi2019QLSAAdiabatic,Liu2021,Costa2022Optimal,Xin2020,Wei2023,Krovi2023Improved,An2025Quantuma}, which are central to numerical simulations. 
Consequently, quantum computing of fluid dynamics (QCFD) has witnessed rapid growth, aiming to harness these capabilities for simulating complex flows~\cite{Givi2020,Givi2021,Succi2023,Meng2025,Tennie2025Quantum}.

Research in QCFD has diversified into several approaches.
One prominent branch involves the quantum lattice Boltzmann method, where extensive efforts have been made to encode collision and transport steps for non-linear fluid dynamics~\cite{Boghosian1998,Yepez2001Lattice,Kocherla2024,Itani2024QALB,Kuya2024,Wawrzyniak2025QALBM,Sanavio2025,Xu2025ImprovedQLBM,Wang2025,Zeng2025,Turro2025,Lee2026}. 
Parallel to these efforts, hybrid quantum-classical frameworks have been developed to offload computationally expensive subroutines to quantum processors~\cite{Gaitan2020,Wang2020,Wang2024,Pfeffer2022,Chen2022,Xue2025,Zhuang2025,Jaksch2023,Bharadwaj2023,Bharadwaj2024,Ingelmann2024,Liu2023,Liu2024,Ye2024,Sedykh2024,Succi2024,Xiao2024,Xiao2025,Yao2025,Zhang2025a,Choi2025,Chen2024,Bharadwaj2025,Wang2025Vortex,Song2025,Alipanah2025}. 
Furthermore, methods based on Hamiltonian simulation have emerged to map fluid dynamics onto the evolution of a quantum system, utilizing the advantages of quantum simulation~\cite{Wright2024,Meng2024a,Meng2023,Meng2024b,Salasnich2024,Lu2024,Zhang2025,Brearley2024,Over2025,Gan2025}. 
Notably, recent studies have extended beyond theoretical algorithm proposals to include proof-of-concept demonstrations on physical quantum processors~\cite{Chen2024,Meng2024a,Bharadwaj2025,Wright2024,Song2025,Alipanah2025,Wang2025Vortex}.

However, the transition to physical hardware introduces the fundamental constraints in the noisy intermediate-scale quantum (NISQ) era~\cite{Preskill2018}. 
NISQ devices possess a limited number of qubits and are susceptible to imperfect gate operations and decoherence~\cite{Cheng2023,Katabarwa2024}. 
As errors accumulate with every operation, deep circuits eventually overwhelm the meaningful signal, limiting algorithms to shallow depths~\cite{Bharti2022Noisy,Meng2025}.
Therefore, any fluid dynamics simulation performed on near-term hardware will inevitably generate results that deviate from the ideal solution due to hardware imperfections.

Although these deviations are widely acknowledged, there is still a lack of physical understanding of how quantum noise specifically alters the flow solution. 
The majority of research has focused on quantifying the magnitude of these errors~\cite{Wright2024,Alipanah2025,Song2025} or implementing error mitigation techniques~\cite{Chen2024,Kuang2025}. 
However, simply treating noise as a stochastic artifact could overlook its potential physical structure. 
For instance, Meng \etal~\cite{Meng2024a} observed that the hardware noise in a superconducting processor produced an energy spectrum resembling the Kolmogorov scaling law of turbulence, suggesting that the quantum noise might mimic physical dissipation at small scales. 
Despite such indications, a systematic characterization is missing. 
It remains an open question whether quantum noise consistently manifests as terms with some physical meanings in the governing equations. 

In this work, we analyze the influence of gate noise on the spectral representation of flow simulation and identify its effective description in the governing partial differential equation (PDE). 
We focus on the one-dimensional (1D) scalar convection using a quantum spectral algorithm, where ideal time advancement modifies only the Fourier phases, effectively isolating the noise-induced changes in spectral magnitudes. 
Modeling the hardware noise as independent Markovian Pauli channels~\cite{Breuer2002}, we derive transition matrices that predict the decay of spectral modes based on the Hamming distance between computational basis indices. 
These predictions are validated against density-matrix simulations and experiments on a superconducting quantum processor. 
Finally, employing a data-driven approach with sparse regression~\cite{Stephany2024PDELearn,Stephany2024WeakPDELearn}, we infer the effective modified PDE governing the noisy evolution. 

The remainder of this paper is organized as follows. 
Section~\ref{sec:theory} outlines the quantum spectral algorithm for scalar convection and derives the transition matrices. 
Section~\ref{sec:method} details the setup for the density-matrix simulations and the experimental implementation on superconducting quantum processor ``Yudu''. 
Section~\ref{sec:result} presents the validation of the noise model and the identification of the effective PDE.  
Finally, Sec.~\ref{sec:summary} summarizes the key findings and discusses potential implications.

\section{Theory}\label{sec:theory}

\subsection{Quantum computing of scalar convection}\label{sec:algorithm}

Scalar convection is a fundamental transport process governing a wide range of phenomena, including combustion and atmosphere transport~\cite{Ren2014Numerical,Yang2023Applications}. 
To analyze how quantum noise influences the simulation, we consider the 1D problem 
\begin{equation}
    \begin{dcases}
      \dfrac{\partial u}{\partial t} + c\dfrac{\partial u}{\partial x} = 0, & (x,t)\in(0,L)\times(0,+\infty),\\
      u(x,0)=f(x),&\\
      u(0,t)=u(L,t),&
    \end{dcases}
    \label{eq:convection}
\end{equation}
with the scalar field $u\lrr{x,t}$, spatial coordinate $x$, and time $t$. 
Its exact solution \(u(x,t)=f(x-ct)\) describes transport of the initial profile $f\lrr{x}$ at constant speed $c$. 
In the Fourier space, the convection modifies only the phases of the Fourier coefficients, leaving their magnitudes invariant. 

\begin{figure*}[htbp]
\centering
  \includegraphics[width=\linewidth]{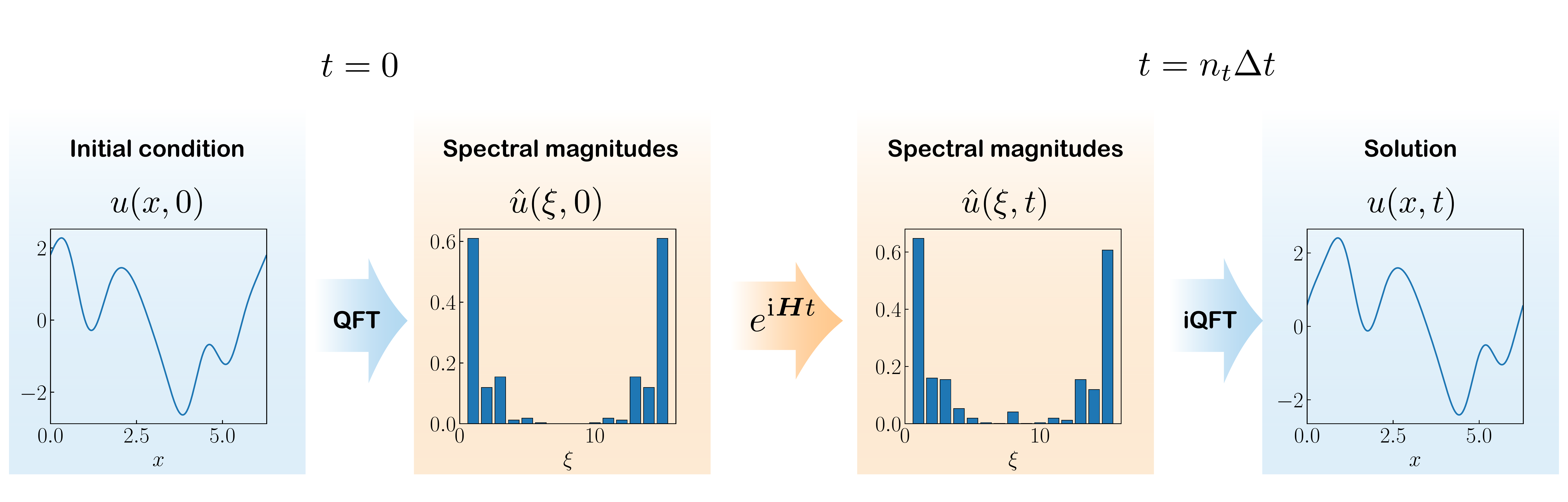}
\caption{Solution procedure for the scalar convection in Eq.~\eqref{eq:convection}. First, the initial condition is encoded into the amplitudes of a quantum state. Next, a QFT maps the state into spectral space. The Hamiltonian simulation in Eq.~\eqref{eq:evo} is then applied for each time step. Finally, an inverse QFT reconstructs a quantum state whose amplitudes represent the solution.
}
\label{fig:solving_steps}
\end{figure*}

The quantum simulation of Eq.~\eqref{eq:convection}, illustrated in Fig.~\ref{fig:solving_steps}, begins by preparing the initial quantum state $\ket{\psi\lrr{t=0}}$, followed by a quantum Fourier transform (QFT). 
In the spectral space, the time evolution over a step $\Delta t$ is~\cite{Lu2024}
\begin{equation}\label{eq:evo}
    \ket{\hat{\psi}(t+\Delta t)}
    =\exp\left(\mathrm{i}c\Delta t\frac{2\pi}{L}D\right)\ket{\hat{\psi}(t)},
\end{equation}
where $\ket{\hat{\psi}(t)}$ denotes the Fourier transform of quantum state $\ket{\psi(t)}$, \(D\) is the diagonal wavenumber operator. 
The operator $D$ admits an efficient decomposition into a sum of Pauli-$Z$ operators~\cite{Meng2023,Meng2024a}, effectively factorizing the time-evolution operator into a tensor product of single-qubit rotations. 
Up to a global phase, the evolution in Eq.~\eqref{eq:evo} for a uniform grid of $N$ points is implemented by a single layer of $R_z$ gates as
\begin{equation}\label{eq:evo_gates}
    e^{ \mathrm{i} c \Delta t\frac{2\pi}{L} D  }
    = e^{ -\mathrm{i} c \Delta t \frac{\pi}{L}I^{\otimes n} }
    \lrs{ \bigotimes_{j=0}^{n-2} R_{z,j} \lrr{ c \Delta t \tfrac{2\pi}{L} 2^{j} }}
    \otimes R_{z,(n-1)} \lrr{ -c \Delta t \tfrac{2\pi}{L} 2^{n-1} },
\end{equation}
where $n=\log_2 N$ is the number of qubits, and $R_{z,j}$ denotes the $R_z$ rotation on the $j$-th qubit, with $j=0,1,\dots,n-1$. 
Finally, an inverse QFT recovers the scalar field in physical space.

In the following section, we provide a theoretical description of how quantum noise affects the Fourier spectrum during this time advancement by Eq.~\eqref{eq:evo_gates}.
Note that this formulation effectively isolates the quantum noise effects. 
Since the evolution operator in Eq.~\eqref{eq:evo_gates} consists exclusively of $R_z$ rotations, it modifies only the relative phases of the computational basis states $\ket{j}$, leaving their probability magnitudes invariant.  
Consequently, any observed deviation in the measured spectral magnitudes can be attributed to the quantum noise, allowing us to characterize the error processes independently of the specific initial condition phases.

\subsection{Modeling the spectral noise}\label{sec:noise_model}

To predict how the measured spectrum deviates from the ideal constant profile, we model the cumulative effect of gate errors using the formalism of open quantum systems. 
We assume that noise acts independently on each qubit at each time step. 
To account for the mixed state nature of the noisy system, we describe the quantum state by a density matrix $\rho$. 
The measurable quantity of interest is the vector $\bm P$, with the $j$-th element $P_j = \expval{\rho}{j}$. 
Physically, since the computational basis encodes the Fourier space, $P_j$ corresponds to the normalized squared amplitude of the $j$-th Fourier mode. 

We first consider the noise effects on a single qubit. 
The evolution of the density matrix under a general noise can be described by the Pauli channels
\begin{equation}
    \rho'=\mathcal{E}\lrr{\rho} = \sum_{k=0}^3 \lambda_k \sigma_k \rho \sigma_k^\dagger,
    \label{eq:rhoprime}
\end{equation}
with the Pauli operators $\sigma_0=I$, $\sigma_1=X$, $\sigma_2=Y$, and $\sigma_3=Z$, and the coefficients satisfying $\sum_{k=0}^3 \lambda_k = 1$. 
Since the computational basis is measured, only the bit-flipping terms $X$ and $Y$ affect measured probabilities, while phase errors $Z$ leave the diagonal elements of $\rho$ invariant. 
Consequently, the relation $\boldsymbol{P}' = \boldsymbol{M}_1 \boldsymbol{P}$ between the input $\bm P$ and the noisy output $\bm P'$ is governed by a transition matrix 
\begin{equation}\label{eq:M_1}    
    \boldsymbol{M}_1 = \begin{pmatrix}
        1 - p/2 & p/2 \\
        p/2 & 1 - p/2
    \end{pmatrix},
\end{equation}
where $p$ is a scalar parameter characterizing the effective noise intensity per step. 
This matrix structure is universal for bit-flip, bit-phase flip, and depolarizing channels, as detailed in~\ref{sec:Pauli}. 

For an \(n\)-qubit system, the transition matrix $\boldsymbol{M}_n$ is the tensor product $\boldsymbol{M}_1^{\otimes n}$ of the single-qubit matrices. 
The probability of a transition between any two states depends on the number of bits flips required to map one to the other. 
If the binary representations of indices $i$ and $j$ differ by a Hamming distance \(d(i,j)\)~\cite{Hamming1950}, the corresponding matrix element is
\begin{equation}\label{eq:M_n}
    M_n[i,j] = (p/2)^{d(i,j)}(1-p/2)^{n-d(i,j)}.
\end{equation}
This transition matrix reveals a hierarchical structure in the spectral noise. 
The spectral modes separated by a small Hamming distance are more likely to mix than distant ones. 
Extending this to a simulation of $l$ time steps, the evolution $\boldsymbol{P}' = \boldsymbol{M}_n^l \boldsymbol{P}$ is described by $l$ successive applications of the transition matrix. 
The repeated application leads to an accumulated effective error probability
\begin{equation}\label{eq:pl}
    p(l) = 1 - \lrr{ 1 - p }^l.
\end{equation}
Substituting Eq.~\eqref{eq:pl} into Eq.~\eqref{eq:M_n} yields the prediction for the noisy spectral evolution after $l$ steps
\begin{equation}\label{eq:Ml_n}
    M_n^l[i,j] = 
    \left[{\frac{1 - \lrr{ 1 - p }^l}{2}}\right]^{d(i,j)}
    \left[{\frac{1 + \lrr{ 1 - p }^l}{2}}\right]^{n-d(i,j)}.
\end{equation}

The transition matrix $M_n^l$ given by Eq.~\eqref{eq:Ml_n} serves as an illustration of noise effects on the spectrum. 
It reveals that the spectral distortion is not a simple random diffusion process. 
Instead, the spectrum relaxes toward a uniform distribution following a hierarchical structure governed by the Hamming distance between the wavenumbers.

\section{Methods of simulation and experiment}\label{sec:method}

\begin{figure*}[htbp]
\centering
  \includegraphics[width=\linewidth]{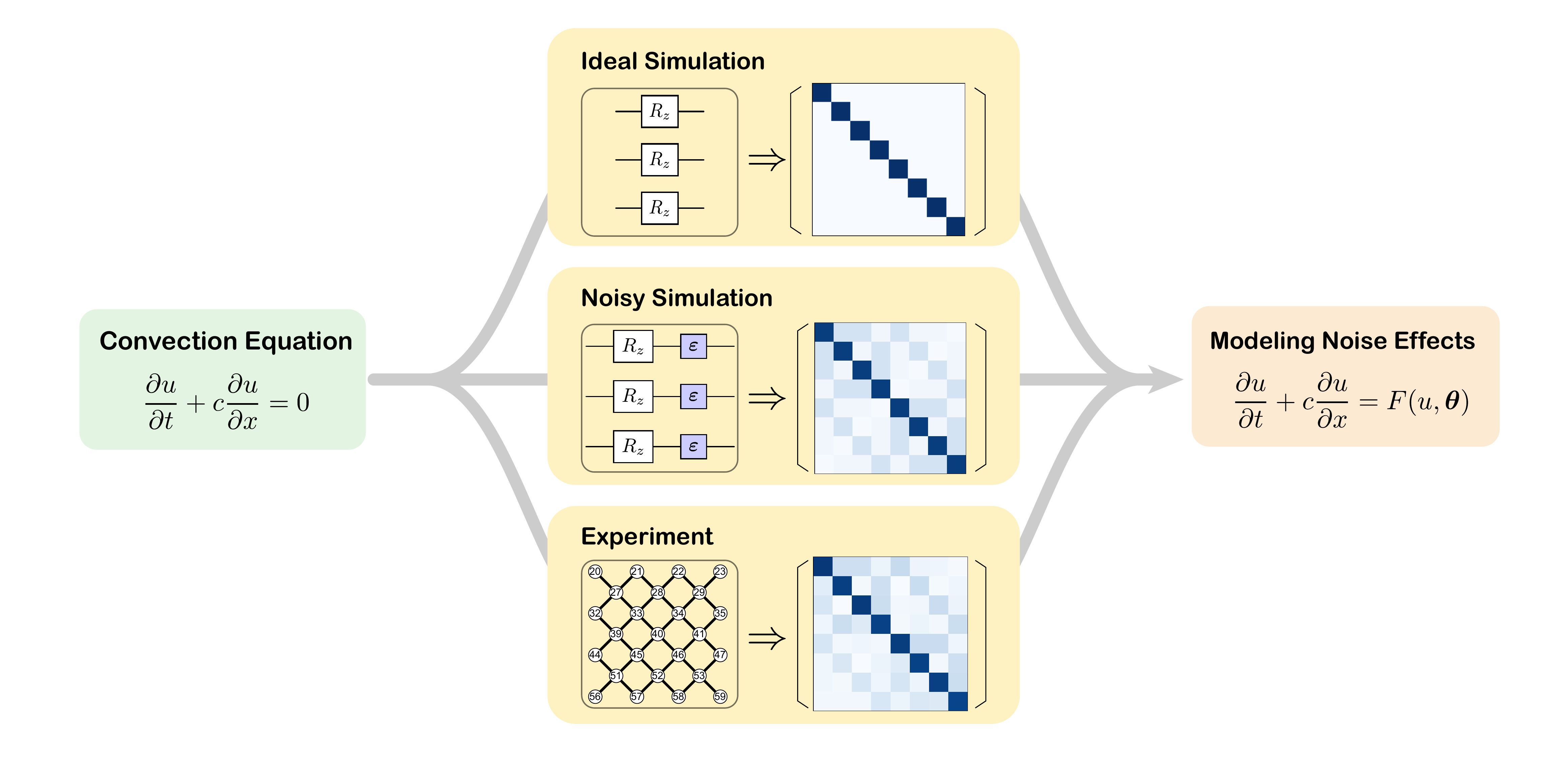}
\caption{
Schematic of different methods for scalar convection: ideal and noisy density-matrix simulations on a classical computer, and experiment on a superconducting processor. 
}
\label{fig:methodology}
\end{figure*}

We investigate the effects of quantum noise on scalar convection simulation using a composite approach. 
As illustrated in Fig.~\ref{fig:methodology}, we conducted ideal state-vector simulations to establish a ground truth. 
In parallel, we characterized noise artifacts using both density-matrix simulations on a classical computer and experiments on the superconducting quantum computer. 
In the three methods, the scalar convection in Eq.~\eqref{eq:convection} was solved with a convection speed $c=1$ in the periodic domain $x \in \lrs{0, 2\pi}$. 
Unless otherwise specified, the spatial domain was discretized using $N=256$ grid points, corresponding to $n=8$ qubits, and the time evolution was computed with a step size of $\Delta t = 0.02$.

\subsection{Quantum simulator}\label{sec:sim}

We used the QPanda framework~\cite{dou2022qpanda,Zou2025} to perform numerical simulations of the quantum circuit corresponding to Eq.~\eqref{eq:evo_gates}. 
Ideal noiseless solutions were computed using the state-vector simulation. 
To study noise effects, we performed density-matrix simulations incorporating the Kraus operator formalism~\cite{nielsen2010,Kraus1971}. 
We introduced uncorrelated single-qubit depolarization channels on the $R_z$ gates, applied identically to all qubits. 
The noise intensities of $p=8.3\times10^{-4}$ and $1.6\times10^{-3}$ were examined, with simulations up to 400 time steps. 
The vector $\bm P'$ was then extracted from the diagonal elements of $\rho'$, as defined in Eq.~\eqref{eq:rhoprime}. 

\subsection{Superconducting quantum computer}\label{sec:exp}

We executed the circuits on superconducting quantum processor ``Yudu'', accessed via the Quafu cloud platform~\cite{baqisQuafuSuperconducting}. 
The key technical specifications of the device are listed in Table~\ref{tab:yudu-params}. 
Quantum computation tasks were constructed and submitted in the form of OpenQASM code. 
As the spectral evolution operator in our algorithm consists exclusively of single-qubit $R_z$ rotations, the circuits were transpiled directly onto linearly connected physical qubits. 
To prevent the transpiler from merging consecutive rotations, barrier instructions were inserted between every layer of $R_z$ gates.
This ensures the preservation of the exact circuit depth $l$, allowing for a noise accumulation analysis. 

\begin{table}[htb]
	\centering
	\begin{tabular}{ll}
		\toprule
		Basis gates      & $H,\ R_x,\ R_y,\ R_z,\ CZ$   \\
		Median single-qubit gate error rate  & $1.56\times10^{-3}$  \\
		Median two-qubit gate error rate & $1.25\times10^{-2}$ \\ 
        Median relaxation time $T_1$ ($\mathrm{\mu s}$)    & $48.07$ \\
        Median dephasing time $T_2$ ($\mathrm{\mu s}$)    & $8.108$ \\
          \bottomrule
	\end{tabular}
	\caption{Specifications of superconducting quantum processor ``Yudu'' on the Quafu platform.}
	\label{tab:yudu-params}
\end{table}

We experimentally reconstructed the transition matrix $\boldsymbol{M}_n^l$ by measuring the transfer between computational basis states $\ket{j}$ with $j=0,1,\dots,2^n-1$. 
For a system of size $n$, we iterated over all $2^n$ computational basis states. 
To prepare a specific initial state $\ket{j}$, we applied $R_y\lrr{\pi}$ gates to qubits corresponding to the binary `1's corresponding to $j$. 
We then applied $l$ layers of the evolution circuit and measured the output $\bm P'$. 
This measured distribution forms the $j$-th column of the experimental transition matrix. 
For each task, the average number of shots for each computational basis state exceeded 600 to minimize the statistical bias. 
Furthermore, we applied the readout error mitigation~\cite{Aasen2025Mitigation} to suppress the influence of measurement errors on the gate errors we sought to characterize. 

\section{Results}\label{sec:result}

\subsection{Scalar convection simulations}

\begin{figure}[htbp]
    \centering
    \begin{subfigure}[b]{0.32\textwidth}
        \centering
        \includegraphics[width=\textwidth]{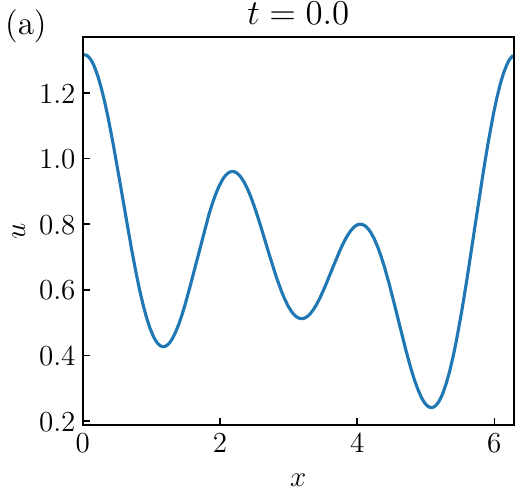}
    \end{subfigure}
    \hfill
    \begin{subfigure}[b]{0.32\textwidth}
        \centering
        \includegraphics[width=\textwidth]{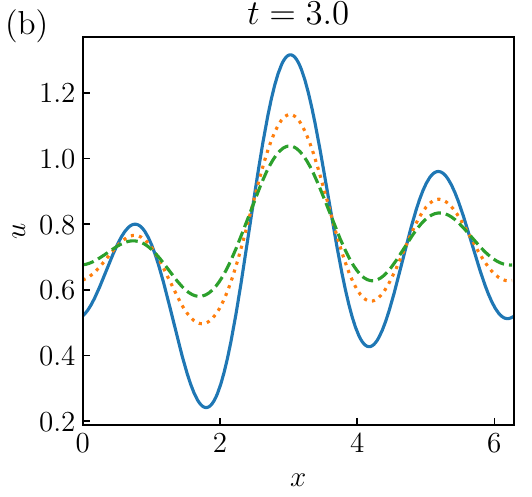}
    \end{subfigure}
    \hfill
    \begin{subfigure}[b]{0.32\textwidth}
        \centering
        \includegraphics[width=\textwidth]{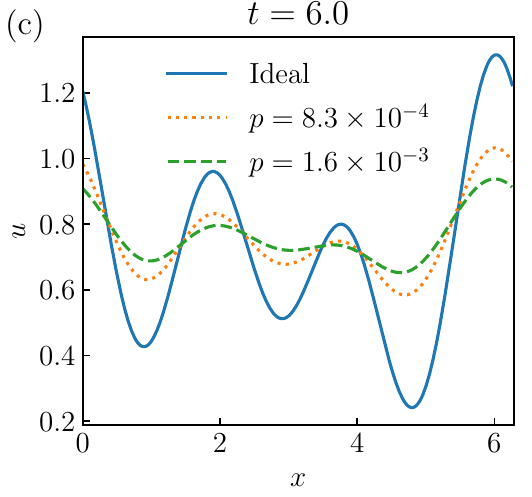}
    \end{subfigure}
    \caption{Evolution of a scalar field undergoing convection in a quantum simulation. 
    (a) Initial profile $u(x,0)=(\cos x + \sin 2x + 2\cos 2x + 3\cos 3x)/10$. 
    (b, c) Snapshots at $t=3.0$ and $t=6.0$ comparing the ideal solution in solid blue lines and density-matrix simulations with $p=8.3\times10^{-4}$ in dotted orange lines and $p=1.6\times10^{-3}$ in dashed green lines. 
    }
    \label{fig:simulation_slns}
\end{figure}

Figure~\ref{fig:simulation_slns} presents the quantum simulation results for the scalar convection problem initialized with a randomly generated profile $u(x,0)=(\cos x + \sin 2x + 2\cos 2x + 3\cos 3x)/10$.
We compare the noisy density-matrix simulations with varying error rates against the baseline ideal simulation.
The ideal quantum simulation result in blue lines propagates the wave at a constant velocity $c=1.0$ along the $x$-axis. 
It preserves the information of the initial state, confirming the unitary nature of the implemented transport operator and the correctness of the noiseless circuit design.

In contrast, the simulations incorporating quantum noise exhibit progressive dissipation and distortion of the scalar field. 
As shown in Figs.~\ref{fig:simulation_slns}(b) and (c), the wave amplitude attenuates significantly over time, with peaks being suppressed and valleys lifting as the system relaxes toward the mean value of the domain, i.e., the maximally mixed state.  
The high-noise simulation ($p=1.6\times10^{-3}$ in dashed green line) exhibits faster attenuation, compared to the lower noise case ($p=8.3\times10^{-4}$ in dotted orange line). 
Beyond simple dissipation, we also observe a distortion in the wave shape. 
The peaks appear slightly shifted relative to the ideal solution.

\subsection{Transition matrices}\label{sec:result_trans_matrix}

\begin{figure}[htb]
	\centering
    \includegraphics[width=\linewidth]{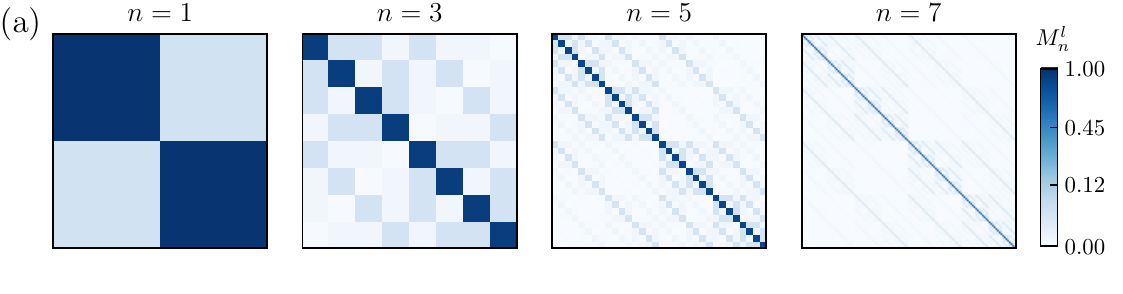}\\
    \includegraphics[width=\linewidth]{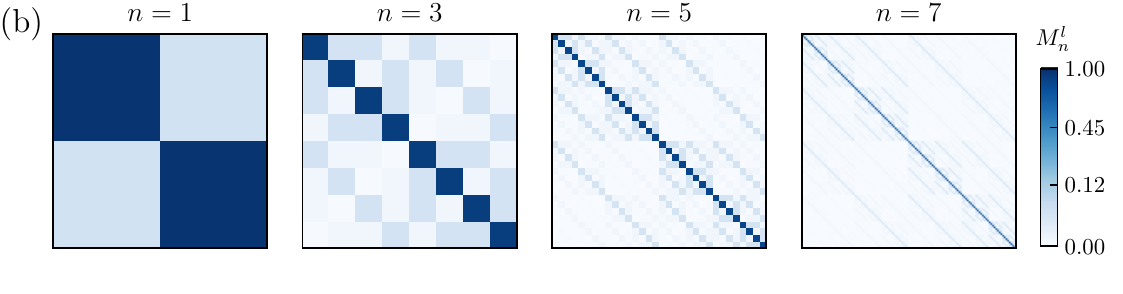}\\
    \includegraphics[width=\linewidth]{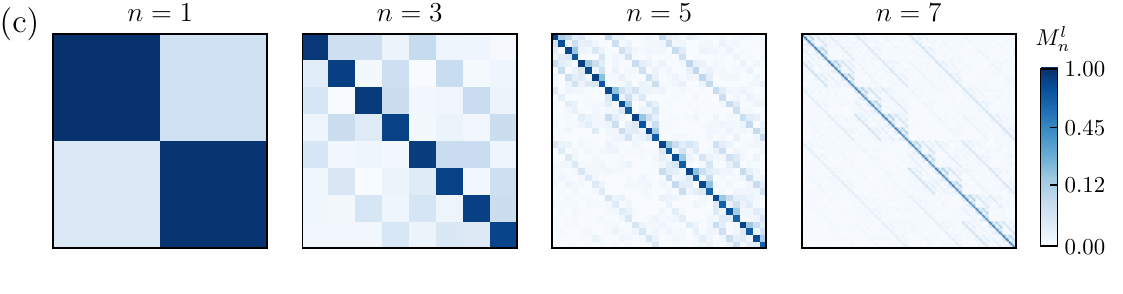}
	\caption{Visualization of the transition matrices for $n=1,3,5,7$ qubits. 
 (a) Theoretical prediction based on Eq.~\eqref{eq:Ml_n} with $p=8.3\times 10^{-4}$. 
 (b) Density-matrix simulation using $p=8.3\times 10^{-4}$ for the depolarizing channel. 
 (c) Experimental reconstruction on superconducting quantum processor ``Yudu''.}
\label{fig:transition_mat}
\end{figure}

To quantitatively characterize the spectral noise, we calculated the transition matrices $\boldsymbol{M}_n^l$ using Eq.~\eqref{eq:Ml_n}, density-matrix simulations, and experiments on the superconducting processor. 
Figure~\ref{fig:transition_mat} presents the results for system sizes $n=1,3,5,7$ at a time step of $l=320$. 
The theoretical and simulated matrices exhibit a self-similar, hierarchical structure. 
This pattern arises from the tensor product $\boldsymbol{M}_n=\boldsymbol{M}_1^{\otimes n}$ of the noise operator, where the magnitude of each element $M_n\lrs{i,j}$ is determined by the Hamming distance $d\lrr{i,j}$ between the row and column indices.  
The agreement between the analytical model in Eq.~\eqref{eq:Ml_n} and the density-matrix simulation validates our derivation.

The experimental matrices obtained from the ``Yudu'' processor shown in Fig.~\ref{fig:transition_mat}(c) largely reproduce this hierarchical structure.
The dominance of the diagonal and the nested off-diagonal blocks validates the hypothesis that the noise acts on the qubits independently to a first approximation. 
However, a closer inspection reveals an asymmetry in the experimental data absent in the model. 

\begin{figure}[htb]
	\centering
	\includegraphics[width=0.45\linewidth]{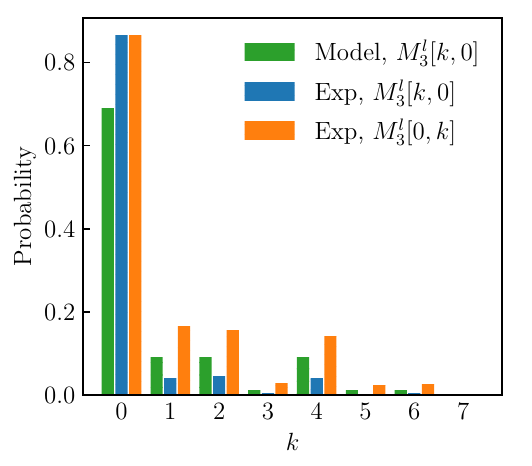}
    \caption{Transition probabilities \( M_n^l[k,0] \) (from ground state $\ket{0}$ to $\ket{k}$, blue hatched bars) and \( M_n^l[0,k] \) (transition from $\ket{k}$ to ground state $\ket{0}$, orange hatched bars) for a three-qubit system after $l=320$ convection layers on superconducting quantum processor ``Yudu'', compared with the symmetric noise model (green dotted bars). 
    }
    \label{fig:bars_Q3}
\end{figure}

To quantify this deviation, Fig.~\ref{fig:bars_Q3} compares the forward and backward transition probabilities between states $\ket{0}$ and $\ket{k}$ for a three-qubit system. 
In the symmetric theoretic model (green bars), these two probabilities equal, $M_3^l[k,0]=M_3^l[0,k]$.
Experimentally, the probability $M_3^l[0,k]$ of transition from $\ket{k}$ to $\ket{0}$ is significantly higher than the probability $M_3^l[k,0]$ of transition from $\ket{0}$ to $\ket{k}$. 
This behavior is a signature of the $T_1$ relaxation~\cite{Carroll2022}, which drives the qubit toward the low-energy state.
This suggests that while the Hamming-distance model captures the primary geometric structure of the noise, detailed quantitative corrections for superconducting hardware requires accounting for physical relaxation.

\begin{figure}[htb]
	\centering

    \begin{subfigure}{0.45\textwidth}
    \centering
    \includegraphics[width=\linewidth]{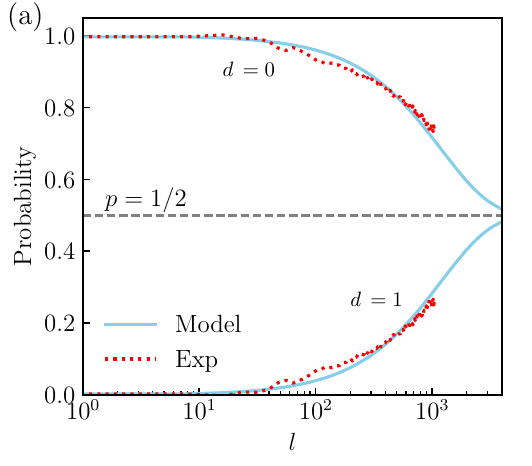}
    \label{fig:eachlevel_evolution_Q1}
    \end{subfigure}
    \hfill
    \begin{subfigure}{0.45\textwidth}
    \centering
    \includegraphics[width=\linewidth]{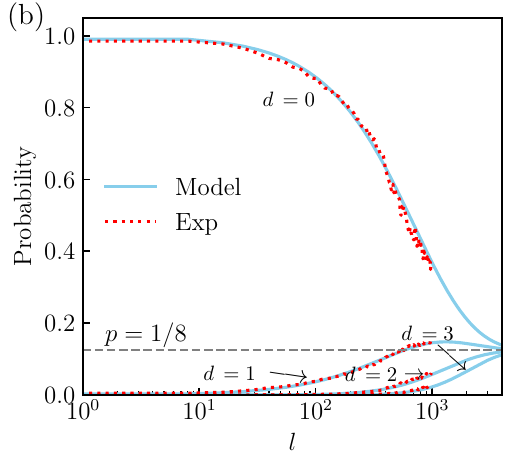}
    \label{fig:eachlevel_evolution_Q3}
    \end{subfigure}
	\caption{Temporal evolution of transition probabilities grouped by Hamming distance $d$. 
    (a) Single-qubit evolution with $d=0, 1$. 
    (b) Three-qubit evolution with $d=0, 1, 2, 3$. 
    Solid lines represent the analytical model in Eq.~\eqref{eq:Ml_n} with fitted $p=8.3\times 10^{-4}$ and dotted lines indicate experimental data from superconducting quantum processor ``Yudu''.}
	\label{fig:eachlevel_evolution}
\end{figure}

Despite this asymmetry, the rate at which the system loses spectral information follows the analytical model.  
Figure~\ref{fig:eachlevel_evolution} tracks the evolution of the transition probabilities grouped by Hamming distance $d$ as a function of the number of time steps $l$.  
The solid lines represent the analytical model in Eq.~\eqref{eq:Ml_n}, using a fitted effective error rate of $p=8.3\times 10^{-4}$.
The probability of maintaining the correct state, $d=0$, decays monotonically from unity. 
At the same time, the probabilities of erroneous transitions, i.e., $d\geq 1$, rise. 
Asymptotically, as $l\rightarrow \infty$, the probability distribution converges to a uniform value $1/2^n$. 
The hierarchical order is preserved throughout the evolution. 
States with smaller Hamming distances from the ideal state always retain higher populations than those further away. 

\subsection{Effective scalar transport}

Having characterized the statistical properties of the noise, we try to identify the PDE that effectively governs the noisy simulated process. 
Using the sparse regression PDE-LEARN~\cite{Stephany2024PDELearn,Stephany2024WeakPDELearn}, detailed in~\ref{sec:PDElearn}, on the simulation data with $p=8.3\times 10^{-4}$, we identified a candidate equation
\begin{equation}\label{eq:pde_identified}
    \frac{\partial u}{\partial t}+c\frac{\partial u}{\partial x} = 0.0471 u + 0.0087 u_{xx} -0.0228 u^2 -0.0021 uu_{xx} + 0.0014 u_x^2 + 0.0013 u_xu_{xx}
\end{equation}
with an unconstrained library of terms.  
Equation~\eqref{eq:pde_identified} suggests the quantum noise introduces a combination of non-linear diffusion and dispersive corrections to the ideal scalar convection.

To isolate the dominant physical mechanisms and improve interpretability, we performed a secondary regression restricted to the terms with the largest contributions, $u$, $u^2$, and $u_{xx}$. 
This approach converged to a simplified reaction-diffusion model
\begin{equation}
    \frac{\partial u}{\partial t}+c\frac{\partial u}{\partial x} = 0.0417u-0.0193u^2+0.0013u_{xx}.
    \label{eq:simple_model}
\end{equation}
It offers a clear physical interpretation of the spectral noise artifacts. 
The diffusion term $u_{xx}$ accounts for the dissipation of high-frequency modes and the smoothing of sharp gradients, and the source terms resemble a reaction mechanism $u(a - bu)$. 
The latter acts as a global amplitude regulator. 
The linear growth and quadratic saturation terms compete to drive the profile toward a stable fixed point, corresponding to the relaxation of the system toward the maximally mixed state. 

\begin{figure}[htbp]
	\centering
    \includegraphics[width=\linewidth]{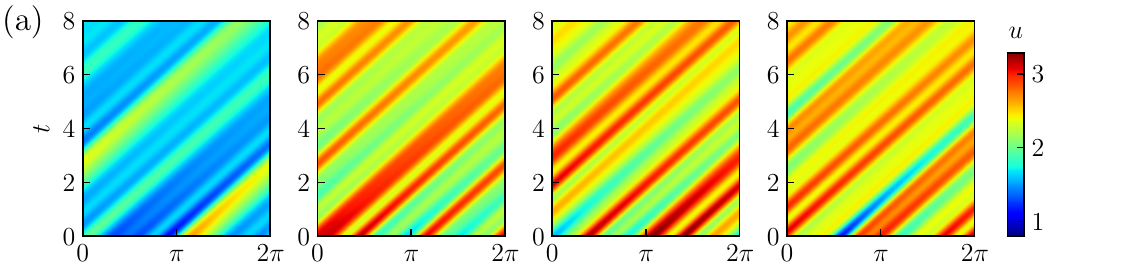} \\
	\includegraphics[width=\linewidth]{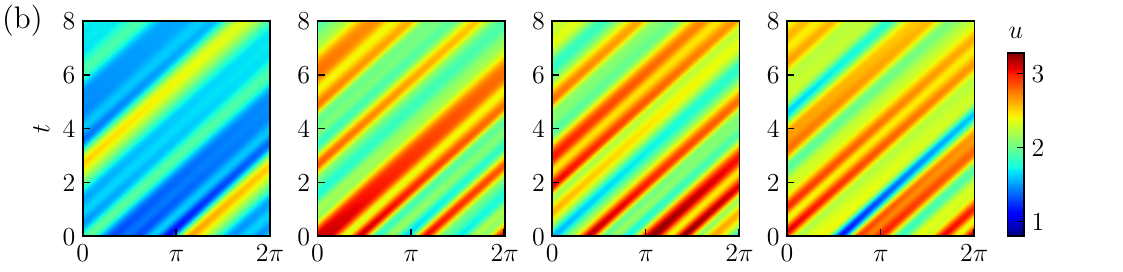} \\
	\includegraphics[width=\linewidth]{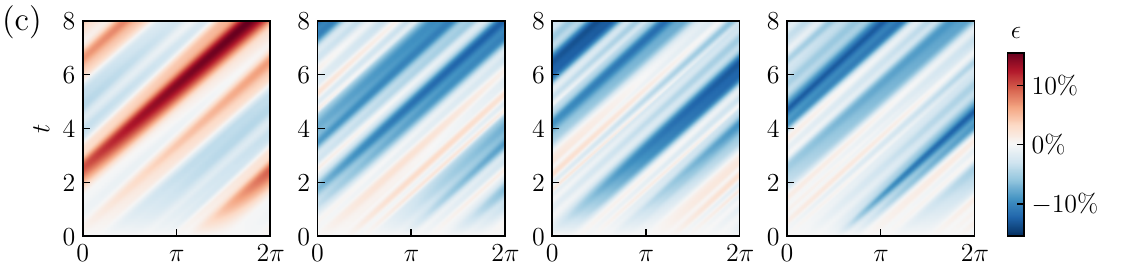}\\
     \includegraphics[width=\linewidth]{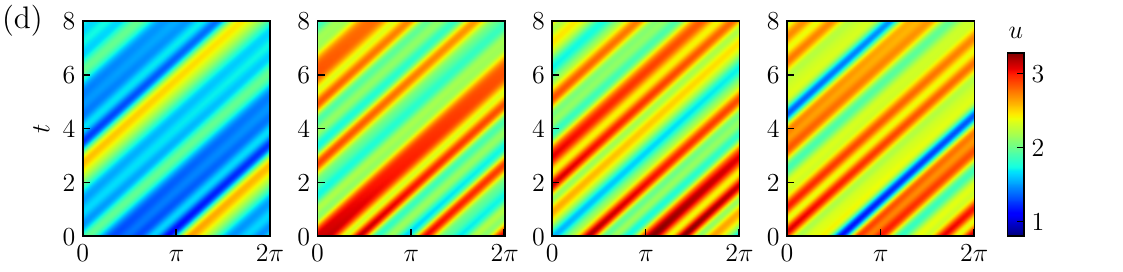} \\
	\includegraphics[width=\linewidth]{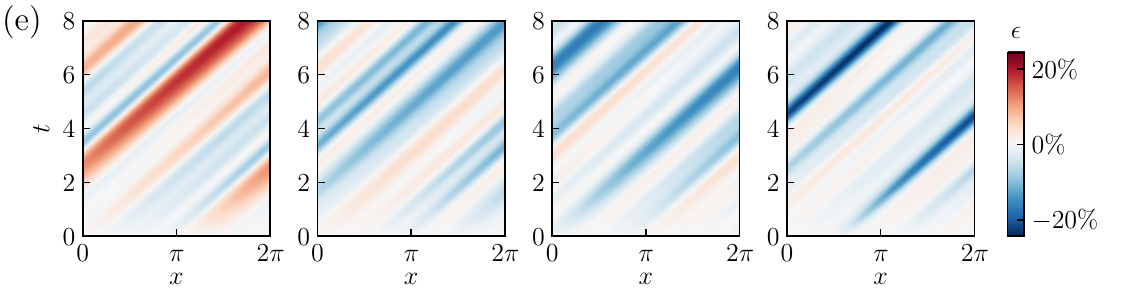}
	\caption{Validation of the effective PDEs discovered via sparse regression.
    (a) Ground truth data from density-matrix simulations. 
    (b) Numerical solution of the full six-term effective PDE in Eq.~\eqref{eq:pde_identified}. 
    (c) Relative error distribution for the six-term model. 
    (d) Numerical solution of the simplified three-term reaction-diffusion model in Eq.~\eqref{eq:simple_model}. 
    (e) Relative error distribution for the simplified three-term model.
    }
	\label{fig:modeling_cases}
\end{figure}

Figures~\ref{fig:modeling_cases} validates these discovered models against independent test datasets initialized with random profiles. 
We compare the ground-truth noisy simulation against the numerical solution of the discovered PDEs in Eqs.~\eqref{eq:pde_identified} and \eqref{eq:simple_model}. 
Both the full six-term model in Eq.~\eqref{eq:pde_identified} and the simplified three-term model in Eq.~\eqref{eq:simple_model} reproduce the dominant features of the noisy quantum simulation results, maintaining a relative error within $\pm 10\%$ versus the majority of the spatio-temporal domain. 
The simplified model achieves comparable predictive accuracy to the full model, confirming that the reaction and diffusion are the primary effective dynamics of the gate noise.

Although the bulk dynamics are well-captured, the error distribution is not uniform. 
As observed in the error plots, deviations accumulate along the characteristic lines, leading to localized divergence in the long-time limit. 
This indicates that although the identified derivative terms successfully model the bulk decay and smoothing, the effective noise operator may contain higher-order or non-local terms necessary to fully regularize the solution over extended integration times. 

\section{Conclusions}\label{sec:summary}

In this work, we analyzed and modeled the effects of gate noise on the quantum simulation of scalar convection. 
Our approach combined theoretical derivation based on open quantum systems, numerical density-matrix simulations, and experiments on a superconducting quantum processor. 
By utilizing a quantum spectral method where the ideal time evolution affects only the phases of the Fourier coefficients, we isolated and characterized how discrete gate errors independently distort the magnitudes of specific spectral modes. 

Assuming independent error channels on the qubits, we derived an analytical model for the noise transition matrix.
We demonstrated that the probability of error-induced transitions is governed by the Hamming distance between the binary representations of the wavenumbers. 
This formulation reveals a recursive spectral decay pattern where the mixing of spectral modes depends on the bit-wise difference of their indices. 
Our model showed quantitative agreement with data from both the quantum simulator and the superconducting hardware, confirming that the recursive Hamming-distance model captures the geometric structure of noise accumulation over varying circuit depths.

We further characterized the effects of noise as additional terms in the scalar transport equation. 
By applying sparse regression to the simulation data, we identified that the spectral errors manifest as effective artificial diffusion and non-linear reaction terms. 
The noise introduces a dissipative term that smooths gradients and a source term that drives the system toward the maximally mixed state. 
These findings suggest that quantum gate noise effects can be modeled as deterministic terms such as artificial viscosity and saturation instead of purely unstructured random perturbations. 

The present study is subject to several limitations that indicate directions for future research. 
The noise modeling is currently restricted to independent, Markovian single-qubit Pauli channels and a 1D scalar convection equation. 
Consequently, theoretical extensions are required to capture the correlations and non-Markovian effects on multi-dimensional flows. 
Furthermore, while the derived PDE qualitatively reproduces the dissipation and smoothing trends, it does not fully match statistical features of the quantum solutions. 
This discrepancy suggests that more expressive model classes, including nonlocal or history-dependent operators, may be needed.

Future work could extend this framework to more realistic noise models and larger-scale flow simulations coupled with advanced error mitigation. 
A promising avenue is to exploit the structured form of the transition matrices to design noise-aware quantum algorithms or to utilize calibrated quantum noise as an effective subgrid model in simulations of complex flows.

\section*{Acknowledgments}

We acknowledge the use of the Quafu cloud quantum computation platform for this work. 
Numerical simulations were carried out on the TianheXY-C supercomputer in Guangzhou, China. 
This work has been supported in part by the National Key R\&D Program of China (Grant No.~2023YFB4502600) and the National Natural Science Foundation of China (Nos.~52306126, 12525201, 12432010, and 12588201).

\appendix

\section{Pauli channels and single-qubit transition matrices}\label{sec:Pauli}

We explicitly derive the single-qubit transition probabilities for standard Pauli noise channels~\cite{nielsen2010}. 
We demonstrate that the bit flip, bit-phase flip, and depolarizing channels all induce an equivalent mixing of the diagonal density matrix elements, parametrized by the error probability $p$. 
Consider a single qubit undergoing a quantum noise channel $\mathcal{E}$. 
The evolution of the density matrix is $\rho'=\mathcal{E}\lrr{\rho}$. 
We focus specifically on the evolution of the diagonal elements
\begin{equation}\label{eq:rhopj}
    \expval{\rho'}{j} = \expval{\mathcal{E}(\rho)}{j},
\end{equation}
with $\ket{j} \in \{\ket{0}, \ket{1}\}$. 
We examine the bit flip channel
\begin{equation}
    \mathcal{E}_{\mathrm{BF}}\lrr{\rho} = (1-\dfrac{p}{2})\rho + \dfrac{p}{2}X\rho X,
\end{equation}
the bit-phase flip channel
\begin{equation}
    \mathcal{E}_{\mathrm{BPF}}\lrr{\rho} = (1-\dfrac{p}{2})\rho + \dfrac{p}{2}Y\rho Y,
\end{equation}
and the depolarizing channel
\begin{equation}
    \mathcal{E}_{\mathrm{DP}}\lrr{\rho} = (1-p)\rho + p\dfrac{I}{2}.
\end{equation}
Substituting these definitions into Eq.~\eqref{eq:rhopj} yields
\begin{equation}\label{eq:tran1}
    \expval{\rho'}{0} = (1-\dfrac{p}{2})\expval{\rho}{0} + \dfrac{p}{2}\expval{\rho}{1}
    \;\mathrm{and}\;
    \expval{\rho'}{1} = (1-\dfrac{p}{2})\expval{\rho}{1} + \dfrac{p}{2}\expval{\rho}{0}.
\end{equation}
It corresponds to the transition matrix in Eq.~\eqref{eq:M_1}.

Meanwhile, the phase flip channel
\begin{equation}
    \mathcal{E}_{\mathrm{PF}}\lrr{\rho} = (1-\dfrac{p}{2})\rho + \dfrac{p}{2}Z\rho Z
\end{equation}
leads to
\begin{equation}\label{eq:tran2}
    \expval{\rho'}{0} = \expval{\rho}{0}
    \;\mathrm{and}\;
    \expval{\rho'}{1} = \expval{\rho}{1}.
\end{equation}
It corresponds to the transition matrix $\bm M_1 = \bm I$.

\section{Data-driven modeling}\label{sec:PDElearn}

We hypothesize that the effective dynamics of the noisy quantum state can be described by a modified continuum equation of the form
\begin{equation}\label{eq:pde2learn}
    \frac{\partial u}{\partial t}+c\frac{\partial u}{\partial x}=F(u,\bm\theta),    
\end{equation}
where $F$ represents an unknown function of the solution and its spatial derivatives, parameterized by the characteristics of the circuit $\bm\theta$.
This approach is motivated by our observation that the noise-induced transition matrix depends solely on the circuit structure and noise channel, effectively acting as valid operator modification independent of the specific fluid state.

To identify $F$, we generated a training corpus using the density-matrix-based quantum simulator described in Sec.~\ref{sec:sim}. 
We constructed 54 distinct datasets corresponding to randomized periodic initial conditions by generating eight Fourier coefficients on an 8D sphere of radius 0.3, and add another three Fourier coefficients on a 3D sphere of radius 0.3 to the first three components of them, as the spectrum of the initial profile.
The simulations utilized a noise intensity of $p=8.3\times10^{-4}$ to match the hardware calibration and evolved for 400 time steps to reach a total time of $T=8.0$.

We employed the PDE-LEARN~\cite{Stephany2024PDELearn,Stephany2024WeakPDELearn} to discover the governing equation from the simulation data. 
The discovery process was designed to isolate the noise-induced deviations by fixing the ideal convection operator $u_t + cu_x$ on the left-hand side of the equation. 
The learning algorithm then selected terms for the right-hand side from a candidate library containing polynomials of $u$ and spatial derivatives up to the fourth order. 
To ensure the resulting model remained physically interpretable, we applied a sparsity threshold of $\varepsilon=5\times10^{-4}$ that masked any term with a coefficient below this value.

The training process was run on the TianheXY-C supercomputer platform with six CPU cores and one NVIDIA A800 GPU. The optimization proceeded through a three-stage protocol consisting of a 300-epoch stabilization phase, a 700-epoch sparsification phase to prune irrelevant terms, and a final 300-epoch fine-tuning phase to optimize the remaining coefficients. 
We validated the discovered PDE by solving it with a classical solver~\cite{comsol} and comparing the results to independent noisy quantum simulations that were not part of the training set.

\bibliography{qNoise}

\end{document}